\documentclass[%
reprint, 
superscriptaddress, 
showpacs,preprintnumbers,
amsmath,amssymb,
aps,
prl,
]{revtex4-2}

\usepackage{graphicx}
\usepackage{dcolumn}
\usepackage{bm}
\usepackage{siunitx}
\usepackage{physics}
\usepackage{xcolor}
\usepackage{soul}
\begin{document}

\preprint{APS/123-QED}

\title{A single length scale rules ballistic aggregation: travels of a droplet train}

\author{Nathan Vani*}
\affiliation{Van der Waals-Zeeman Institute, University of Amsterdam, Science Park 904, Amsterdam, Netherlands
}%
\email{nathan.vani@ens-paris-saclay.fr}

\author{Stefan Kooij}
\affiliation{Van der Waals-Zeeman Institute, University of Amsterdam, Science Park 904, Amsterdam, Netherlands
}%
\author{Aritro Mukherjee}
  \affiliation{ 
Faculty of Physics, University of Duisburg-Essen, Lotharstraße 1, Duisburg, Germany
}%
  \author{Antoine Parrenin}
\affiliation{Van der Waals-Zeeman Institute, University of Amsterdam, Science Park 904, Amsterdam, Netherlands
}%
  \author{Cees J.M. van Rijn}
\affiliation{Van der Waals-Zeeman Institute, University of Amsterdam, Science Park 904, Amsterdam, Netherlands
}%
\author{Daniel Bonn}
\affiliation{Van der Waals-Zeeman Institute, University of Amsterdam, Science Park 904, Amsterdam, Netherlands
}

\date{\today}

\begin{abstract}

Ballistic aggregation is a canonical non-equilibrium process, relevant across scales from granular gases to planetary accretion. Collisions are driven by differences in velocities, building up persistent correlations between neighbors. Here, we provide the first experimental realization of one-dimensional ballistic aggregation in a train of droplets formed by the breakup of a liquid jet. Experiments and simulations confirm the analytically predicted scaling, with the global process shown to be governed by a single length scale. While air drag inverts the sign of neighbor velocity correlations, a 1D ordering constraint protects bulk characteristics of ballistic aggregation such as the scaling exponent and the shape of the large mass tail. More broadly, our results show that Smoluchowski-like mean-field descriptions fail when collisions carry directional memory -- as demonstrated here for jet-generated sprays.

\end{abstract}

\maketitle

Ballistic aggregation describes the dynamics of particles traveling at constant velocity between collisions, merging irreversibly while conserving mass and momentum, in contrast to elastic or diffusion-limited processes. The process is of particular interest to granular gases~\cite{zaburdaev2006free}, the formation of traffic jams~\cite{ben1994kinetics}, thin-films~\cite{brett1988structural}, hailstones~\cite{lozowski1991simulating} and planets~\cite{dominik1997physics}, and, as we argue here, to the coalescence of liquid sprays. The seminal work of Carnevale, Pomeau, and Young (CPY) used scaling arguments to predict that the average particle mass grows as $\langle m \rangle \propto t^{\zeta}$ with $\zeta = 2d/(d+2)$, where $t$ is the time of evolution and $d$ is the dimension~\cite{Carnevale1990}. In any finite dimension $d > 1$, the prediction rests on an invalid mean-field assumption, and thus fails~\cite{Trizac2003}. The ballistic process indeed generates correlations that violate Boltzmann's Stosszahlansatz or molecular-chaos hypothesis, a critical assumption in kinetic theory models which presume that the discrete constituents of the model remain uncorrelated before and after collisions~\cite{leyvraz2003scaling, Krapivsky2010_Art_BBGKY,Stoss_Contemp}.


The one-dimensional case is special as the CPY scaling
holds – it even admits an exact self-similar solution in its
asymptotic regime. Frachebourg first derived it by mapping the dynamics onto a Brownian motion problem with absorbing parabolic boundaries, thereby demonstrating that the solution violates the Stosszahlansatz~\cite{Frachebourg1999, frachebourg2000ballistic}. To date, no experimental confirmation of the scaling has been provided -- nor has Frachebourg's predicted scaling function been confirmed numerically.

\begin{figure}[t!]
  \centering
  \includegraphics[width=\columnwidth]{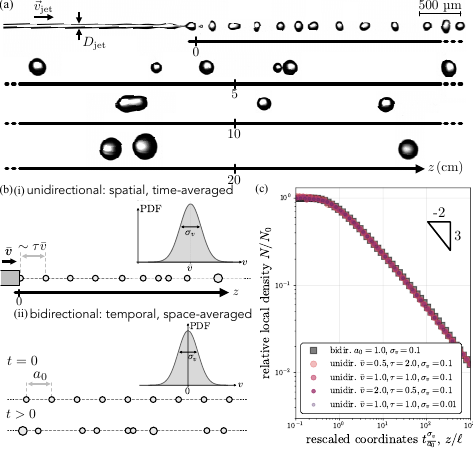}
  \caption{(a) Inherently unstable, a liquid jet breaks up into a train of droplets. As we look further downstream of the train we see ever fewer and larger droplets. Corresponding videos are available as supplementary material. (b) We model the train as a ‘unidirectional' ballistic aggregation process (i) of average speed $\bar{v}$ with a standard deviation $\sigma_v$ at $z = 0$ and an emission period $\tau$. Measurements are localized in space. The system is equivalent to a ‘bidirectional' one (ii) with a zero average speed in which measurements are localized in time -- as long as the initial standard deviations in velocity match and the initial particle spacing $a_0$ matches $\tau\bar{v}$. (c) Collapse of local density $N/N_0$ as a function of the rescaled coordinates $t\sigma_v/a_0$ and $z/\ell$ for several simulation parameters.
  }
  \label{fig:intro_expe_solver}
\end{figure}

We argue that trains of droplets are a good candidate to do so. In Fig.~\ref{fig:intro_expe_solver}a, we show a liquid jet of diameter $D_{\mathrm{jet}}$ traveling at speed $v_{\mathrm{jet}}$ breaking up into a nearly monodisperse train of droplets due to the propagation of capillary waves~\cite{Eggers2008}. Looking further downstream along the axis $z$, differences in velocities lead to merging events with ever bigger and sparser droplets. Transverse displacements remain well below the droplet size, keeping the train effectively one-dimensional. Depending on flow rate, satellite droplets can be produced at break up~\cite{van2010breakup} -- but they are of no consequence here as their velocity spread is large enough for them to quickly merge into main droplets~\cite{van2014velocity}. We consider here jets of diameters 32 to 180 $\mu$m, using high-speed imagery to track individual particles at different positions downstream. Experimental methods are detailed in the End Matter.

In this Letter, we first show that trains of droplets can be modeled through ballistic aggregation with downstream distance playing the role of time in the comoving frame. Using scaling arguments, we demonstrate that the system is fully ruled by a single length scale, a prediction we confirm both numerically and experimentally. We then focus on fine-grained properties of the system, i.e. mass distributions and correlation coefficients, showing good agreement between theory and simulation. Drag shifts the experiments away from pure ballistic aggregation though its signature, such as the shape of the large mass tail, remains. Surprisingly, drag inverts the sign of neighbor correlations while leaving the growth exponent untouched.


Fig.~\ref{fig:intro_expe_solver}b (i) shows a schematic of a kinematic model for a train of droplets. We consider point-like particles, as the spacing between particles is large enough relative to their radius and oscillation -- an assumption which clearly holds after a few merging events. During these events, both mass and momentum are conserved. We consider particles monodisperse in mass. The system is characterized by three parameters: the period of emission $\tau$, the initial average velocity $\bar{v}$ and its initial spread $\sigma_v$, with the system varying in space along $z$. In Fig.~\ref{fig:intro_expe_solver}b (ii), we draw the canonical ballistic aggregation model, set by an initial particle spacing $a_0$, a zero average velocity, and a spread $\sigma_v$, with the system varying in time along $t$. Using event-driven solvers, we show that the systems are statistically equivalent upon the mapping $t = z/\bar{v}$ (Fig.~\ref{fig:end_solver} of End Matter).

Dimensional analysis indicates that the system depends on a single dimensionless number. Scaling arguments à la CPY recover it: consider a clump of particles of the canonical system in its asymptotic regime at time $t$, made of the merging of $N$ particles of mass $m_0$ with its mass $m = Nm_0$, and its momentum $P$. To reach this size, the clump must have swept the typical distance $a_0N$ over time $t$, traveling at a typical velocity $\delta P/m$ relative to the mean, such that $a_0N = t\delta P/m$. Since the clump results from the aggregation of $N$
particles whose original velocities were independently drawn, momentum conservation lets the Central Limit Theorem set its typical momentum fluctuation: $\delta P \sim \sqrt{N} (m_0\sigma_v)$, with the average velocity conserved. This is precisely where the CPY prediction breaks down for $d > 1$: velocity correlations in the system invalidate the CLT, whereas the geometrically imposed ordering allows the calculation in 1D. Matching the equations, we find: $m/m_0 = (t\sigma_v/a_0)^{2/3}$, which, applied to the unidirectional geometry, gives:

\begin{equation}
    \frac{\langle m \rangle }{m_0} \sim \left(\frac{\sigma_v}{\tau\bar{v}^2} z\right)^{2/3},
\label{eq:mass_collapse}
\end{equation}

\noindent matching both CPY and Frachebourg scaling predictions with the added identification of the single length scale $\ell \equiv \tau\bar{v}^2/\sigma_v$. This length corresponds to the typical distance two particles must travel before meeting. Fig.~\ref{fig:intro_expe_solver}c shows the near-perfect collapse, across various system parameters, of the simulated local density. Note that the initial velocity distribution was picked here to be normal to reproduce experimental conditions, but a similar collapse can be found with other distributions of varying skewness and kurtosis.

Experimentally, the Rayleigh-Plateau instability predicts both the initial particle distance to average $4.5D_{\mathrm{jet}}$ and the initial droplet diameter as $1.89D_{\mathrm{jet}}$. The average initial velocity is straightforwardly conserved from $v_{\mathrm{jet}}$. The experimental value of $\ell$ thus depends only on $D_{\mathrm{jet}}$ and the ratio $\sigma_v/v_{\mathrm{jet}}$. We conducted experiments across three jet diameters (32–180 \SI{}{\micro\meter}) and velocities from 2–120 \SI{}{\meter/\second}, in five combinations, with Reynolds numbers spanning 320 to 3800, Ohnesorge numbers nearly constant at 0.015-0.021 as liquid properties are not varied, and air Weber numbers spanning 0.03 to 7.7, reaching a regime in which aerodynamic effects might affect breakup for the smaller, faster jet. Initial droplet velocities follow narrow normal distributions, with relative standard deviation varying from 1 to 6\% (Table~\ref{tab:dimensionless_numbers}). 

\begin{figure}[h]
  \centering
  \includegraphics[width=\columnwidth]{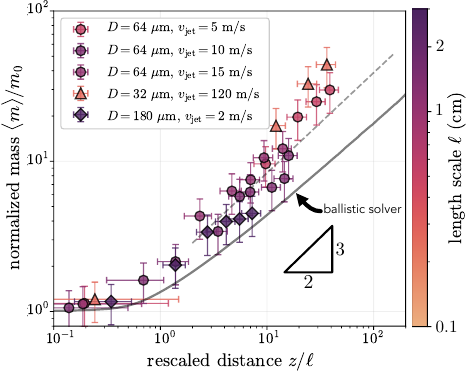}
  \caption{Evolution of the average mass normalized by Rayleigh-Plateau prediction of the initial droplet mass $m_0$ as a function of the rescaled distance $z/\ell$ with $\ell$ shown in the colorbar, for several jet sizes and velocities. The full gray line is the prediction from the ballistic solver. The dashed line indicates a $2/3$ slope, whereas individual fits of the slope average $0.67 \pm 0.08$.}
  \label{fig:mass_rescaled}
\end{figure}

In Fig.~\ref{fig:mass_rescaled}, we show the good collapse of experimentally measured $\langle m \rangle$ following Eq.~\eqref{eq:mass_collapse}, each point averaging between 10,000 and 4,000,000 droplets. Since errors are partly systematic, we conduct individual slope fits for each dataset for $z/\ell > 1$ (see End Matter for error estimation). We find an average exponent of $0.67 \pm 0.08$, consistent with the predicted $2/3$ growth law and clearly sublinear. The spread is relatively large -- possibly due to the limited number of $z$ positions probed per condition and the lack of a fully established asymptotic regime ($z/\ell < 100$). Prediction from the ballistic solver is plotted along the data: in the growth regime ($z/\ell > 1$) a larger prefactor is observed experimentally, of the order of 2 rather than 1 from the model. We attribute this discrepancy to (i) the effect of air drag, as discussed thereafter, (ii) finite radius-effects, especially important near breakup due to inertial oscillations. We expect the latter to increase the effective rescaled coordinate and the former to modify the dynamics. Drag is indeed size-dependent, with deceleration of each droplet scaling inversely with its size. Drag thus increases the likelihood of merging of small droplets with larger ones, effectively removing smaller droplets from the system and increasing the average size. We now turn to fine-grained properties in which the effects of the drag are more clearly visible.

\begin{figure}[h]
  \centering
  \includegraphics[width=\columnwidth]{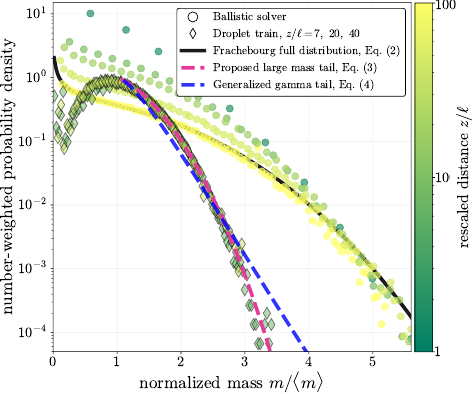}
  \caption{Number-weighted probability density function of the normalized mass from simulations (circles) and experiments (diamonds). At large $z/\ell$, the former agrees with the Frachebourg distribution (Eq.~\eqref{eq:frach_full}, black line). Experimental data is narrower due to drag. We fit its tail successfully with the Frachebourg-derived proposed expression (Eq.~\ref{eq:large-mass-frach}, pink dashed line) and without success with a Smoluchowski-derived generalized gamma (Eq.~\ref{eq:gen_gamma}, blue dashed line) -- both are fitted with a free amplitude prefactor. Experimental data correspond to trains of droplets originating from a jet of diameter $D_{\mathrm{jet}}=64\mu$m, at either $v_{\mathrm{jet}}=$ 10 m/s ($z/\ell=7$) or $v_{\mathrm{jet}}=$ 5 m/s ($z/\ell=20$, 40).}  \label{fig:distribs}
\end{figure}

In Fig.~\ref{fig:distribs}, we plot the number-weighted probability density function of the normalized mass in the system as predicted by the ballistic solver (circles). At large $z/\ell$, the solver converges towards Frachebourg's analytical solution:

\begin{equation}
    p(x) = \dfrac{x\,I(x)\,H(x)}{\int_0^\infty \,x'\,I(x')\,H(x')\,\dd x'}, \quad x = m/\langle m\rangle,
    \label{eq:frach_full}
\end{equation}

\noindent where $I(x) = \sum_{k=1}^{\infty} e^{-\omega_k x}$ and $H(x) = \frac{1}{2\pi i}\int_{-i\infty}^{+i\infty}\frac{e^{-xw}}{\mathrm{Ai}^2(w)}\dd w$ ~\cite{Frachebourg1999, frachebourg2000ballistic}, with $\mathrm{Ai}$ the Airy function. Note that when number-weighted, the analytical mass distribution diverges as $1/\sqrt{m}$ for small masses. To our knowledge, this is the first numerical validation of the scaling function derived by Frachebourg.

Experimental data from droplet trains (diamonds) do not agree with prediction from ballistic aggregation as the distributions are narrower. This observation is in accordance with the proposed drag mechanism rather than pre-asymptotic transients since the shapes collapse over several $z/\ell$. It is still possible to extract a signature of ballistic aggregation in the shape of the large mass tail, which we expect to remain unaffected by the merging of small droplets. We thus propose the following form, for the large mass tail derived from Frachebourg doubly-asymptotic expression~\cite{frachebourg2000ballistic}:

\begin{equation}
        p(x) = \frac{x^{5/2}}{I_0\, \lambda^{5/2}} \exp\left(-\frac{\lvert \omega_1 \rvert x}{\lambda} - \frac{x^3}{12\lambda^3}\right), \quad x = m/\langle m\rangle,
        \label{eq:large-mass-frach}
\end{equation}

\noindent where $\omega_1 \simeq -2.34$ is the first negative zero of the Airy function, $\lambda$ a fitting parameter, and $I_0 = \lambda \int_0^\infty u^{5/2}\exp\left(\omega_1 u - \frac{u^3}{12}\right)\dd u$ a normalization factor.  In contrast, the distribution of droplet size sprays is usually predicted by Smoluchowski-type models \cite{marmottant2004fragmentation,villermaux2004ligament,Kooij2018} which result in a generalized gamma distribution for droplet diameters. Expressed in terms of mass (derivation in End Matter), the distribution is given by:

\begin{equation}
    p(x) = \frac{(\nu c)^{\nu}}{3\,\Gamma(\nu)}\; x^{\frac{\nu - 3}{3}}\; \exp(-\nu c\, x^{1/3}), \quad x = m/\langle m\rangle,
    \label{eq:gen_gamma}
\end{equation}

\noindent where $\Gamma$ is the gamma function, $\nu$ is a fitting parameter, and $c^3 = (\nu+1)(\nu+2)/\nu^2$. We fit the large mass tail ($m/\langle m \rangle > 1$) with both expressions, adding a free amplitude to both. We find Eq.~\ref{eq:large-mass-frach} to describe well the shape of the large mass tail (pink line, $\lambda = 10$), whereas Eq.~\ref{eq:gen_gamma} fails (blue line, $\nu = 65$). The generalized gamma stretched exponential decay ($\propto e^{-x^{1/3}}$) is too heavy-tailed to reproduce the experimental sharp cutoff contrarily to the super-exponential ($\propto e^{-x^{3}}$) suppression predicted by ballistic aggregation. Since coalescence is driven by velocity-correlations, mean-field models fail to capture the essential physics of the system while ballistic aggregation successfully reproduces the large mass tail. 

\begin{figure*}[t!]
    \includegraphics[width = \textwidth]{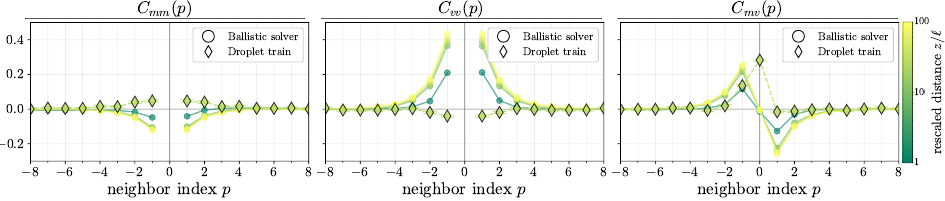}
    \caption{Correlation coefficients from the simulations (circles) and from experiments (diamonds) $C_{mm}$, $C_{vv}$, $C_{mv}$ for various neighbor index $p$, with $p > 0$ denoting droplets farther downstream. Experimental data correspond to $400.000$ droplets recorded for $D_{\mathrm{jet}}=64\mu$m, $v_{\mathrm{jet}}=5$m/s at $z=15$ cm, corresponding to $z/\ell=29$. Trivial values of $C_{mm}(0) = C_{vv}(0) = 1$ are omitted. Further experimental measurements of correlation are shown in the SM.}
    \label{fig:correlations}
\end{figure*}

We finally note that the characterization of the large mass tail was only made possible thanks to the high-precision, single droplet sizing method used. More conventional spray measurement methods, such as diffraction-based ones, add multiplicative noise whose heavier tail then dominates the observed distribution. It is notable that mean-field models appear here best at predicting data of poor quality in correlated systems.

We conclude by computing the correlations in simulations (circles) and experiments (diamonds) in Fig.~\ref{fig:correlations}, expressed in mass-mass, velocity-velocity and mass-velocity as a function of the neighbor index $p$ which increases along the $z$ direction. We consider Pearson linear correlation coefficients, e.g. defined for $C_{mm}$ as:

\begin{equation}
    C_{mm}(p) \equiv \frac{\sum_i (m_i - \langle m\rangle)(m_{i+p} - \langle m\rangle)}{\sum_i (m_i - \langle m\rangle)^2}.
    \label{eq:def_corr}
\end{equation}

\noindent Estimating experimental error is of particular importance here as coefficients are generally low. Such low values of correlation coefficients are sometimes used to justify mean-field models~\cite{vledouts2015fragmentation} -- but without proper error estimation, they do not demonstrate a lack of correlation.

In both simulations and experiments, we uncover non-zero correlations for close neighbors ($ p < 4$), confirming that mean-field approaches are not suited to model the system. The experimental coefficients were often found to have lower absolute values, as expected given experimental noise. We first consider the signs of the coefficients for the ballistic solver. We find $C_{mm}$ to be strictly negative though weak: large droplets get larger by depleting the region around them, decreasing the probability of finding another large particle nearby. $C_{vv}$ is strictly positive with a larger value: as originally pointed out by CPY, particles tend to travel in packs of similar velocity -- a survivorship bias. $C_{mv}$ is antisymmetric as the process itself is spatially symmetric. Antisymmetry alone forces $C_{mv}(0) = 0$: physically, this reflects that the velocity of a growing particle relaxes without bias toward $\bar{v}$.


Experimentally, $C_{mv}(0) =  0.26 \pm 0.01$, a strong indicator of differential drag as mass and velocity are positively correlated. $C_{mv}$ is also not antisymmetric, with a relatively large positive value of $C_{mv}(1)$, potentially a signature of the slipstream effect~\cite{parrenin2024effect}. The signs of $C_{mm}$ and $C_{vv}$ are flipped with regard to the ballistic solver. Intriguingly, $C_{mm} > 0$ and $C_{vv} < 0$ indicate that droplets travel preferably with neighbors of similar size but of different velocities. A full description of these correlation patterns, in drag-free simulations and in drag-contaminated experiments,  is beyond the scope of this Letter. In both cases, non-zero velocity correlations indicate the violation of the Stosszahlansatz, invalidating mean-field Smoluchowski dynamics. It is remarkable that despite different correlation behaviors, they both follow a similar scaling, are ruled by the same length scale, and display a similarly shaped large mass tail.

Our results show that trains of droplets allow the probing of ballistic aggregation, and conversely that the framework of ballistic aggregation is well suited to model jet-generated sprays. Along the way, we provided experimental and numerical validations of analytical results and demonstrated how the length scale $\ell$ governs the process. Drag influences both the global mass distribution and the clustering structure while preserving bulk properties. Droplet trains warrant further study: quantitative, drag-dependent models of size distributions are still lacking as mean-field models fail due to velocity correlations. By controlling ambient air pressure~\cite{parrenin2026ambient}, one could probe pure ballistic aggregation through drag-free trains of droplets. The regime $z/\ell < 1$ also happens to be the one relevant to inkjet printing~\cite{lohse2022fundamental}. Furthermore, we expect the scaling argument to extend to other strongly directional sprays, relevant to drug-delivery applications.

Our results are of interest beyond sprays and fundamental questions of ballistic aggregation alone. For example, the proposed length scale $\ell$ collapses data from a recent study on the self-assembly of bacteria from Sintès~\textit{et al.}~\cite{sintes2026swimming}. The presented framework should also naturally extend to other soft matter systems governed by ballistic merging, such as emulsion trains~\cite{dewandre2020microfluidic}, post-breakup particles from gas~\cite{riviere2025breakup} or liquid~\cite{thievenaz2022fragmentation} ligaments, oceanic bubble streams~\cite{rehder2002enhanced} and microorganism groups emerging from active instabilities~\cite{eisenmann2025pure}. More broadly, our results suggest investigating velocity correlations wherever Smoluchowski-type mean-field models are commonly applied -- for example in open problems related to planet-formation dynamics~\cite{Planet_formation_article}.\\

\paragraph{Acknowledgments --} D.B. and N.V. thank Maria Tătulea-Codrean for enjoyable conversations on ballistic aggregation. N.V. acknowledges the help of Samantha Kucher on the programming of solvers. A.M. acknowledges the guidance and support of S.M.P. Devi. This research has been funded by the Dutch Research Council NWO, IPP grant “Innovative Nanotech Sprays”, and by ERC grant “RayleighSprays” ID:101142159.

\paragraph{AI usage --} Claude Sonnet 4.5 was used as a coding assistant in the development of the ballistic solvers and the generation of figures.

\paragraph{Data availability --} Python scripts for ballistic solvers and for estimation of Frachebourg's scaling function, as well as processed experimental data is available at https://github.com/NathanVani/BallisticAggreg. Raw data is available upon reasonable request.

\bibliographystyle{unsrt}  
\bibliography{references}

\clearpage
\section{End Matter}

\begin{figure*}
    \includegraphics[width = 0.9\textwidth]{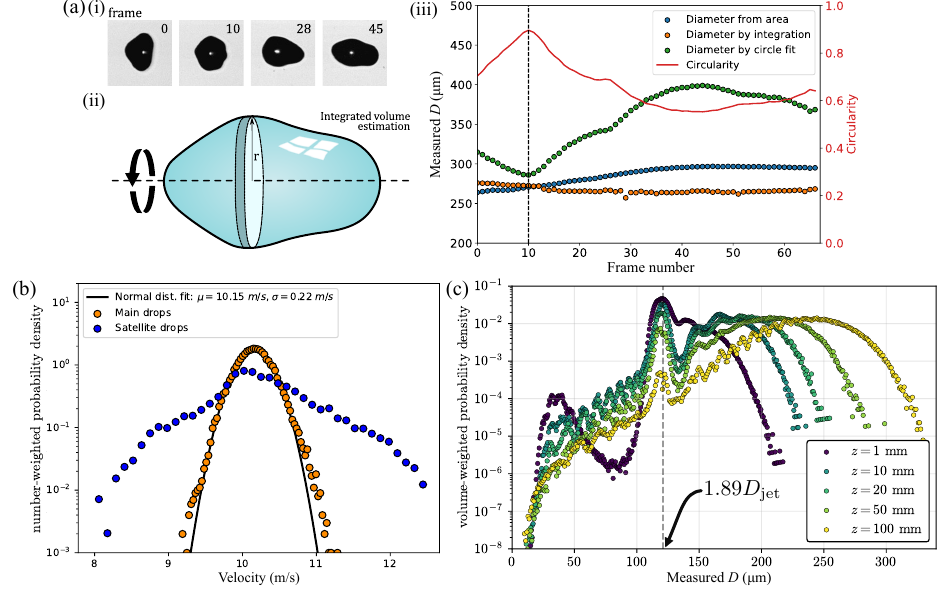}
    \caption{(a) Visual volume-integration method. (i) Time sequence of an oscillating droplet resulting from a merging event. (ii) Integration of the volume of the droplet under assumption of axisymmetry. (iii) Apparent diameter of the droplet inferred from area-based projection, circular fits, and axisymmetric volume integration. Area-based and circular fit show variations of approximately 30 and 15\%, whereas volume-integration varies only by about 8\%. All methods agree when the droplet matches a spherical shape. (b) Velocity distributions of main (orange) and satellite (blue) droplets near jet breakup, for $D_\mathrm{jet} =$ \SI{64}{\micro\meter} and $v_\mathrm{jet} =$ \SI{10}{\meter\per\second}. The main droplets follow a narrow normal distributions while satellite droplets display a broader distribution. (c) Evolution of the volume-weighted droplet diameter distribution at several distances from the nozzle $z$, for $D_\mathrm{jet} =$ \SI{64}{\micro\meter} and $v_\mathrm{jet} =$ \SI{10}{\meter\per\second}. Dashed line corresponds to the Rayleigh-Plateau predicted initial droplet size.}
    \label{fig:end_matter_expe}
\end{figure*}

\paragraph{Experimental methods.} We generate jets by pumping liquid at an imposed flow rate using an HPLC pump (Waters~515) through either a custom micro-fabricated spray nozzle (diameter of 32 and \SI{64}{\micro\meter}, MedSpray) or a disposable 27G commercial straight needle (diameter of \SI{180}{\micro\meter}).

All experiments reported here are performed with water at room temperature ($T \approx \SI{293}{K}$, $\rho = \SI{998}{kg\,m^{-3}}$, $\gamma = \SI{72}{mN\,m^{-1}}$, $\mu = \SI{1}{mPa\,s}$). Small amounts of table salt are added to suppress self-charging of the spray. Transverse displacements of the droplets due to air perturbation were limited by using a protective transparent box set around the trajectory of the train. The jet is pointed downward, though given the droplet size and speed, gravity plays no significant role in the experiments.

Jet breakup and droplet trains evolution are recorded at set distance from the nozzle using a high-speed Phantom TMX~7510 camera fitted with either a Sigma Macro 105mm or a 25mm Ultra Macro lens. Droplets are backlit using an LED light source.

A key element of our analysis is a volume-integration droplet-sizing method that is robust to the strong shape distortions induced by coalescence-driven oscillations, as illustrated in Fig.~\ref{fig:end_matter_expe}a. By reconstructing the droplet volume from the axisymmetric silhouette rather than relying on projected area or circular fits, this approach yields more accurate estimations of droplet volumes. The improved fidelity is essential for precisely resolving the size of droplets for distributions such as the large mass tail in Fig.~\ref{fig:distribs} which would be obscured with other sizing methods.

Immediately following jet breakup, droplets exhibit a narrow velocity distribution centered near the jet exit velocity. Figure~\ref{fig:end_matter_expe}b shows the velocity density of droplets just after breakup of a jet of diameter $D_{\mathrm{jet}} =$ \SI{64}{\micro\meter} and speed $v_{\mathrm{jet}} =$ \SI{10}{\meter \per \second}. The main droplets are well characterized by a narrow normal distribution centered on $v_{\mathrm{jet}}$ with a relative velocity spread of only $\sigma_v / v_{\mathrm{jet}} = 2\%$. In contrast, satellite droplets -- small secondary droplets produced between main drops -- exhibit a significantly broader velocity distribution. Satellites therefore merge rapidly with main droplets, so that they play no role in the subsequent dynamics. Rare collision events such as droplets bouncing or producing satellite while merging were not observed during these experiments -- consistently with the considered Weber numbers. Fig.~\ref{fig:end_matter_expe}c shows the evolution of the volume-weighted diameter distribution at different distances from the nozzle. Parameter sets used for the experiments discussed in the Letter are presented in Table~\ref{tab:dimensionless_numbers}.

\begin{table}[h!]
    \centering
    \begin{tabular}{ccccccc}
    \hline
    $D_{\mathrm{jet}}$ (\SI{}{\micro\meter}) & $v_{\mathrm{jet}}$ (\SI{}{\meter\per\second}) & $\sigma_v/v_{\mathrm{jet}}$ (\%) & $\ell$ (cm) & $\mathrm{Re}$ & $\mathrm{Oh}$ & $\mathrm{We}_{\mathrm{air}}$ \\
    \hline
    32  & 120 & 1.1 & 0.12 & 3800 & 0.021 & 7.68 \\
    64  & 5   & 5.6 & 0.51 & 320  & 0.015 & 0.03 \\
    64  & 10  & 2.0 & 1.07 & 640  & 0.015 & 0.11 \\
    64  & 15  & 2.7 & 1.44 & 960  & 0.015 & 0.24 \\
    180 & 2   & 2.8 & 2.89 & 360  & 0.009 & 0.01 \\
    \hline
    \end{tabular}
    \caption{Measured relative velocity spread, computed length scale and dimensionless numbers characterizing the five experimental jet conditions: the length scale $\ell = \tau\bar{v}^2/\sigma_v$, with $\tau = 4.5D_{\mathrm{jet}}/v_{\mathrm{jet}}$ set by the Rayleigh--Plateau initial spacing, the Reynolds number $\mathrm{Re} = \rho v_{\mathrm{jet}} D_{\mathrm{jet}}/\mu$, the Ohnesorge number $\mathrm{Oh} = \mu/\sqrt{\rho \sigma D_{\mathrm{jet}}}$, and the aerodynamic Weber number $\mathrm{We}_{\mathrm{air}} = \rho_{\mathrm{air}} v_{\mathrm{jet}}^2 D_{\mathrm{jet}}/\sigma$}.
    \label{tab:dimensionless_numbers}
\end{table}

\paragraph{Estimation of errors.} Errors in Fig.~\ref{fig:mass_rescaled} combine several sources: uncertainty in calibration, and in droplet volume estimation (dominant for the mass measurement, of the order of 8\%), in $z$ position, and in $v_{\mathrm{jet}}/\sigma_v$ and $D_{\mathrm{jet}}$. Part of these errors is systematic for a single data set, hence the consideration of individual fits. A Monte Carlo method is used to propagate the errors.

For the correlation coefficients, velocity series are detrended to remove slow drift from pump fluctuations (window size of 500 measurements chosen through a convergence test). Because successive droplets are correlated, the effective number of independent samples $N_{\mathrm{eff}}$ is slightly smaller than the raw droplet count; we estimate it from the integrated autocorrelation time of each series. The standard error on each Pearson coefficient then follows the usual expression $\left(1 - C(p)^2\right)/\sqrt{N_{\mathrm{eff}}}$.

\paragraph{Gamma distribution expressed in mass.} The mean-field approach commonly used for sprays predicts the distribution in diameter to follow~\cite{marmottant2004fragmentation}:

\begin{equation}
p_x(x) = \frac{\nu^\nu}{\Gamma(\nu)}\, x^{\nu-1}\, e^{-\nu x}, \qquad x = D/\langle D \rangle.
\end{equation}

\noindent The associated mass $m$ can be expressed as $m = \frac{\pi}{6}\rho \langle d^3 \rangle x^3$. We then introduce the variable $y \equiv m/\langle m \rangle = x^3/\langle x^3 \rangle$. Since $x$ follows a gamma distribution, we have:

\begin{equation}
    \langle x^3\rangle = \frac{(\nu+1)(\nu+2)}{\nu^2} \equiv c^3.
\end{equation}

\noindent Thus $x = cy^{1/3}$. The Jacobian corresponds to $dx/d y = cy^{-2/3}/3$. We finally get the probability in rescaled mass $y$ as:

\begin{equation}
    p(y) = \frac{(\nu c)^{\nu}}{3\,\Gamma(\nu)}\; y^{\,\nu/3 - 1}\; e^{-\nu c\, y^{1/3}},
\end{equation}

\noindent a generalized gamma function of shape factor $\nu/3$, power $1/3$, and scale $(\nu c)^{-3}$.

\begin{figure}[h]
  \centering
  \includegraphics[width=\columnwidth]{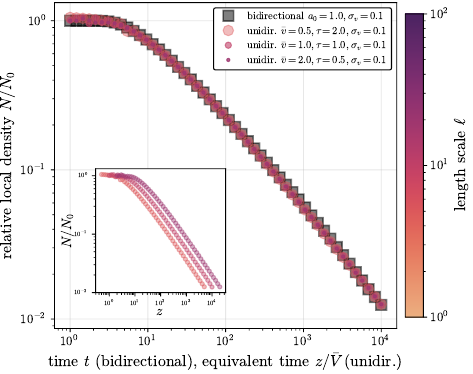}
  \caption{Evolution of the relative local density $N/N_0$ as a function of time $t$ (bidirectional) or equivalent time $z/\bar{v}$. Inset shows the original unidirectional curves.}
  \label{fig:end_solver}
\end{figure}

\paragraph{Event-driven solvers.} To probe both transitory and asymptotic regimes of ballistic aggregation, we programmed two event-driven solvers for the bidirectional and unidirectional systems, tracking point-like Lagrangian particles. Since particles move at constant velocity, this approach is exact and efficient. Furthermore, we have checked that the scaling in $\ell$ holds for other initial velocity distributions (uniform, log-normal). Codes are made publicly available. Fig.~\ref{fig:end_solver} shows the equivalency between the two systems in terms of particle density. Note that for the bidirectional system, the reference density corresponds to the initial system-wide density: $a_0^{-1}$, while for the unidirectional one it is $(\tau\bar{v})^{-1}$. Measurements are made over a time (bidirectional) or space (unidirectional) window small compared to the characteristic time or length scale.

\clearpage

\end{document}